\documentclass[twocolumn,showpacs,superscriptaddress,amsmath,amssymb]{revtex4}
\usepackage{graphics}
\usepackage{graphicx} 
\usepackage{dcolumn} 
\usepackage{amssymb}

\usepackage{graphicx}
\usepackage{dcolumn}
\usepackage{bm}
\def\bs{$\backslash$} 
 
\def\bd{\begin{document}} \def\ed{\end{document}}
\def\bmp{\begin{minipage}} \def\emp{\end{minipage}}
\def\bcc{\begin{center}} \def\ecc{\end{center}}     \def\npg{\newpage}
\def\beq{\begin{equation}} \def\eeq{\end{equation}} \def\hph{\hphantom}
\def\be{\begin{equation}} \def\ee{\end{equation}} \def\r#1{$^{[#1]}$}
\def\n{\noindent} \def\ni{\noindent} \def\pa{\parindent}
\def\hs{\hskip} \def\vs{\vskip} \def\hf{\hfill} \def\ej{\vfill\eject}
\def\cl{\centerline} \def\ob{\obeylines}  \def\ls{\leftskip}
\def\underbar#1{$\setbox0=\hbox{#1} \dp0=1.5pt \mathsurround=0pt
   \underline{\box0}$}   \def\ub{\underbar}    \def\ul{\underline}
\def\f{\left} \def\g{\right} \def\e{{\rm e}} \def\o{\over} \def\d{{\rm d}}
\def\vf{\varphi} \def\pl{\partial} \def\cov{{\rm cov}} \def\ch{{\rm ch}}
\def\la{\langle} \def\ra{\rangle} \def\EE{e$^+$e$^-$} \def\pt{p_{\rm T}}
\def\dt{\delta}
\def\bitz{\begin{itemize}} \def\eitz{\end{itemize}}
\def\btbl{\begin{tabular}} \def\etbl{\end{tabular}}
\def\btbb{\begin{tabbing}} \def\etbb{\end{tabbing}}
\def\beqar{\begin{eqnarray}} \def\eeqar{\end{eqnarray}}
\def\\{\hfill\break} \def\dit{\item{-}} \def\i{\item}
\def\bbb{\begin{thebibliography}{9} \itemsep=-1mm}
\def\ebb{\end{thebibliography}} \def\bb{\bibitem}
\def\bpic{\begin{picture}(260,240)} \def\epic{\end{picture}}
\def\akgt{\noindent{Acknowledgements}}
\def\fgn{\noindent{\bf\large\bf figure captions}}
\def\lan{\langle}
\def\ran{\rangle}
\def\p{\pi}
\def\ifmath#1{\relax\ifmmode #1\else $#1$\fi}%
\def\rc{\ifmath{{\mathrm{c}}}}
\def\cut{\ifmath{{\mathrm{cut}}}}
\def\rF{\ifmath{{\mathrm{F}}}}
\def\rK{\ifmath{{\mathrm{K}}}}
\def\rp{\ifmath{{\mathrm{p}}}}
\def\rt{\ifmath{{\mathrm{t}}}}
\def\LAB{\ifmath{{\mathrm{LAB}}}}
\def\cut{\ifmath{{\mathrm{cut}}}}
\def\beq{\begin{equation}}
\def\eeq{\end{equation}}
\def\us{^{(s)}}  \def\bea{\begin{eqnarray}} \def\eea{\end{eqnarray}}
\def\nbr{\nonumber} \def\e{\eta} \def\D{\Delta}
\def\r{\rho}  \def\unln{\underline}
\newcommand{\cinst}[2]{$^{\mathrm{#1}}$~#2\par}
\newcommand{\crefi}[1]{$^{\mathrm{#1}}$}
\newcommand{\crefii}[2]{$^{\mathrm{#1,#2}}$}
\newcommand{\crefiii}[3]{$^{\mathrm{#1,#2,#3}}$}
\newcommand{\HRule}{\rule{0.5\linewidth}{0.5mm}}
\usepackage{color}
\newcommand{\Blue}[1]{\textcolor[named]{Blue}{#1}}
\newcommand{\blue}[1]{\textcolor[named]{Blue}{#1}}
\newcommand{\Red}[1]{\textcolor[named]{Red}{#1}}
\newcommand{\red}[1]{\textcolor[named]{Red}{#1}}
\newcommand{\violet}[1]{\textcolor[named]{Violet}{#1}}
\newcommand{\brown}[1]{\textcolor[named]{Brown}{#1}}
\newcommand{\green}[1]{\textcolor[named]{Green}{#1}}
\newcommand{\magenta}[1]{\textcolor[named]{Magenta}{#1}}

\def\yw{y_{\rm w}}  \def\ew{\eta_{\rm w}}  \def\de{\delta\eta} \def\bs{B_s} \def\ssnn{\sqrt {s_{NN}}}

\def\Journal#1#2#3#4{{#1} {\bf #2}, #3 (#4)}

\begin{document}


\title{Longitudinal scaling property of the charge balance function
 in Au + Au collisions\\ at $\ssnn =200$ GeV}

\affiliation{Argonne National Laboratory, Argonne, Illinois 60439,
USA} \affiliation{University of Birmingham, Birmingham, United
Kingdom} \affiliation{Brookhaven National Laboratory, Upton, New
York 11973, USA} \affiliation{University of California, Berkeley,
California 94720, USA} \affiliation{University of California, Davis,
California 95616, USA} \affiliation{University of California, Los
Angeles, California 90095, USA} \affiliation{Universidade Estadual
de Campinas, Sao Paulo, Brazil} \affiliation{University of Illinois
at Chicago, Chicago, Illinois 60607, USA} \affiliation{Creighton
University, Omaha, Nebraska 68178, USA} \affiliation{Czech Technical
University in Prague, FNSPE, Prague, 115 19, Czech Republic}
\affiliation{Nuclear Physics Institute AS CR, 250 68
\v{R}e\v{z}/Prague, Czech Republic} \affiliation{University of
Frankfurt, Frankfurt, Germany} \affiliation{Institute of Physics,
Bhubaneswar 751005, India} \affiliation{Indian Institute of
Technology, Mumbai, India} \affiliation{Indiana University,
Bloomington, Indiana 47408, USA} \affiliation{Alikhanov Institute
for Theoretical and Experimental Physics, Moscow, Russia}
\affiliation{University of Jammu, Jammu 180001, India}
\affiliation{Joint Institute for Nuclear Research, Dubna, 141 980,
Russia} \affiliation{Kent State University, Kent, Ohio 44242, USA}
\affiliation{University of Kentucky, Lexington, Kentucky,
40506-0055, USA} \affiliation{Institute of Modern Physics, Lanzhou,
China} \affiliation{Lawrence Berkeley National Laboratory, Berkeley,
California 94720, USA} \affiliation{Massachusetts Institute of
Technology, Cambridge, MA 02139-4307, USA}
\affiliation{Max-Planck-Institut f\"ur Physik, Munich, Germany}
\affiliation{Michigan State University, East Lansing, Michigan
48824, USA} \affiliation{Moscow Engineering Physics Institute,
Moscow Russia} \affiliation{City College of New York, New York City,
New York 10031, USA} \affiliation{NIKHEF and Utrecht University,
Amsterdam, The Netherlands} \affiliation{Ohio State University,
Columbus, Ohio 43210, USA} \affiliation{Old Dominion University,
Norfolk, VA, 23529, USA} \affiliation{Panjab University, Chandigarh
160014, India} \affiliation{Pennsylvania State University,
University Park, Pennsylvania 16802, USA} \affiliation{Institute of
High Energy Physics, Protvino, Russia} \affiliation{Purdue
University, West Lafayette, Indiana 47907, USA} \affiliation{Pusan
National University, Pusan, Republic of Korea}
\affiliation{University of Rajasthan, Jaipur 302004, India}
\affiliation{Rice University, Houston, Texas 77251, USA}
\affiliation{Universidade de Sao Paulo, Sao Paulo, Brazil}
\affiliation{University of Science \& Technology of China, Hefei
230026, China} \affiliation{Shandong University, Jinan, Shandong
250100, China} \affiliation{Shanghai Institute of Applied Physics,
Shanghai 201800, China} \affiliation{SUBATECH, Nantes, France}
\affiliation{Texas A\&M University, College Station, Texas 77843,
USA} \affiliation{University of Texas, Austin, Texas 78712, USA}
\affiliation{Tsinghua University, Beijing 100084, China}
\affiliation{United States Naval Academy, Annapolis, MD 21402, USA}
\affiliation{Valparaiso University, Valparaiso, Indiana 46383, USA}
\affiliation{Variable Energy Cyclotron Centre, Kolkata 700064,
India} \affiliation{Warsaw University of Technology, Warsaw, Poland}
\affiliation{University of Washington, Seattle, Washington 98195,
USA} \affiliation{Wayne State University, Detroit, Michigan 48201,
USA} \affiliation{Institute of Particle Physics, CCNU (HZNU), Wuhan
430079, China} \affiliation{Yale University, New Haven, Connecticut
06520, USA} \affiliation{University of Zagreb, Zagreb, HR-10002,
Croatia}

\author{B.~I.~Abelev}\affiliation{University of Illinois at Chicago, Chicago, Illinois 60607, USA}
\author{M.~M.~Aggarwal}\affiliation{Panjab University, Chandigarh 160014, India}
\author{Z.~Ahammed}\affiliation{Variable Energy Cyclotron Centre, Kolkata 700064, India}
\author{A.~V.~Alakhverdyants}\affiliation{Joint Institute for Nuclear Research, Dubna, 141 980, Russia}
\author{I.~Alekseev~~}\affiliation{Alikhanov Institute for Theoretical and Experimental Physics, Moscow, Russia}
\author{B.~D.~Anderson}\affiliation{Kent State University, Kent, Ohio 44242, USA}
\author{D.~Arkhipkin}\affiliation{Brookhaven National Laboratory, Upton, New York 11973, USA}
\author{G.~S.~Averichev}\affiliation{Joint Institute for Nuclear Research, Dubna, 141 980, Russia}
\author{J.~Balewski}\affiliation{Massachusetts Institute of Technology, Cambridge, MA 02139-4307, USA}
\author{L.~S.~Barnby}\affiliation{University of Birmingham, Birmingham, United Kingdom}
% ...

\collaboration{STAR Collaboration}\noaffiliation


\date{\today}

\begin{abstract}
We present measurements of the charge balance function, from the
charged particles, for diverse pseudorapidity and transverse
momentum ranges in Au + Au collisions at $\ssnn =200$ GeV using the
STAR detector at RHIC. We observe that the balance function is
boost-invariant within the pseudorapidity coverage [-1.3, 1.3]. The
balance function properly scaled by the width of the observed
pseudorapidity window does not depend on the position or size of the
pseudorapidity window. This scaling property also holds for
particles in different transverse momentum ranges. In addition, we
find that the width of the balance function decreases monotonically
with increasing transverse momentum for all centrality classes.
\end{abstract}



\maketitle

Particle production in elementary collisions at high energy is
constrained by conservation laws. Electric charge conservation, in
particular, constrains the balance of charged particles produced in
a collision. The electric charge balance function (BF) is an
observable specifically designed to measure the balance, and thereby
provide insight into the particle production processes in elementary
collisions at high energy~\cite{oldbf1}. It has been used in
hadron-hadron, lepton-hadron, and e$^+$e$^-$ collisions to study
hadronization schemes~\cite{oldbf1,lh,e+e-}. The BF has recently
gained particular interest in clocking hadronization in relativistic
heavy-ion collisions, where a new state of matter - the quark-gluon
plasma (QGP) - would be formed. The formation of QGP will allow a
partonic charge diffusion in longitudinal phase space, and would
lead to a widening of the charge balance function~\cite{bf1}.

The BF is defined in terms of a combination of four different
conditional densities of charged hadrons~\cite{oldbf1}. It measures
how the net charge at any point of the phase space is rearranged if
the charge at a selected point changes. Projected on to the
pseudorapidity difference $\de =\eta_1-\eta_2$  of two charged
particles in a given pseudorapidity window $\ew$, the BF
becomes~\cite{bf1,star130}
\begin{eqnarray}
 B(\de|\ew) &=& \frac{1}{2}\left[
 \frac{\la n_{+-}(\de,\ew)\ra-\la n_{++}(\de,\ew)\ra}{\la
 n_+(\ew)\ra}\right.
 \nonumber \\
 & &\quad \left.+\frac{\la n_{-+}(\de,\ew)\ra-\la n_{--}(\de,\ew)\ra}{\la
n_-(\ew)\ra}\right]
\end{eqnarray}
where $\la n_+(\ew)\ra $ and $\la n_-(\ew)\ra $ are respectively the
event averaged number of measured positively and negatively charged
particles. $\la n_{+-}(\de,\ew)\ra = \la n_{-+}(\de,\ew)\ra $ is the
event averaged number of pairs of particles with opposite charges
separated by pseudorapidity $\de$. $\la n_{++}(\de,\ew)\ra$ and $\la
n_{-+}(\de,\ew)\ra $ are defined correspondingly for pairs of
positively and negatively charged particles, respectively. The
charge balance function is a differential combination of all
possible charge correlations. Its integral over rapidity space is
related to measures of charge fluctuation~\cite{bf2}.

Measurements of the BF in relativistic heavy-ion collisions have
been reported by several experiments~\cite{star130,na49,phenix}.
However, these experiments feature significant difference of
acceptance in pseudorapidity and transverse momentum. Comparison of
results from these experiments is thus only qualitative. A
quantitative comparative analysis of these results requires a better
understanding of the BF dependence on pseudorapidity and momentum
acceptance~\cite{star130,na49,na22bf,tom1}.

This BF dependence has been studied in $\pi^{+}\rp$ and $\rK^{+}\rp$
collisions at 250 GeV/$c$ incident beam momentum by a fixed target
experiment with large acceptance~\cite{na22bf}. In those collisions,
the BF is found to be invariant under longitudinal boost over the
whole rapidity range of produced particles ($-5 < y < 5$), i.e., the
ratio of $B(\dt y|\yw)$ to $\left(1-\dt y/|\yw|\right)$ is
independent of the observed window, $|\yw|$, and corresponds to the
BF of the whole rapidity range.

The aim of the analysis presented in this paper is to verify whether
the boost invariance observed in elementary collisions is also
present in Au + Au collisions at $\ssnn = 200$ GeV. To this end, we
first study the BF using equal size pseudorapidity windows spanning
various pseudorapidity ranges within a relatively wide
pseudorapidity coverage of the STAR Time Projection Chamber (TPC).
Given that the shape of the BF depends by definition on the width of
the pseudorapidity acceptance, we next scale the measured BF by an
acceptance factor determined by the width of the observed
pseudorapidity window. The scaled balance function, $\bs(\de)$, is
defined as
\begin{eqnarray}
   \bs(\de)= \frac{B(\de|\ew)}{1-\frac{\de}{|\ew|}}
  \end{eqnarray}
where $\de$ is the particle separation in pseudorapidity, and
$|\ew|$ represents the size of pseudorapidity window. This scaled BF
shows how the balance function extends with the widening of the
pseudorapidity window.

\begin{figure*}
\includegraphics[width=6.6in]{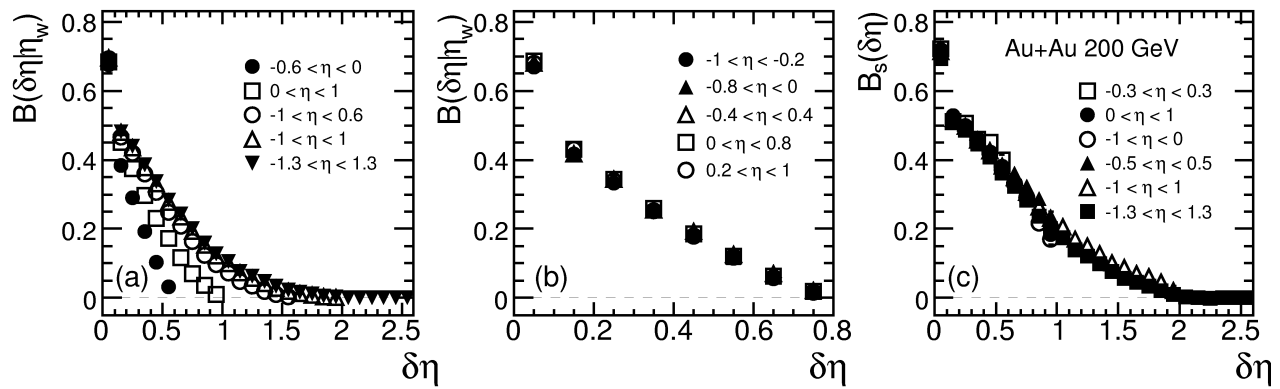}
\caption{\label{Fig. 1} (a) Balance functions in five pseudorapidity
windows of different width; (b) Balance functions observed at five
different positions of pseudorapidity windows with $|\ew| = 0.8$;
(c) Scaled balance function, $\bs(\de)$, obtained for various
pseudorapidity window widths and positions. The data are from
$0-80\%$ Au+Au collisions at 200 GeV and the particle $\pt$ range is
$0.15<\pt<2$ GeV/c. Statistical errors are smaller than the symbol
sizes. Systematic errors are of the order of $5\%$.}
\end{figure*}

We further explore the scaling property of the BF in different
ranges of transverse momentum $\pt$. The $\pt$ of final state
particles is suggested to characterize their emission proper-time
$\tau$~\cite{rudy, bmuller, hama1}. Particles with different $\pt$
may be produced at different stages of the evolution after the
collision. This relation between $\pt$ and $\tau$ has been assumed
in hydrodynamic models~\cite{hydro}, which qualitatively describe
the data of $\pt$ dependence of anisotropic collective
flow~\cite{v2}. Examining the $\pt$ dependence of the scaling of the
BF will provide an additional experimental test of this assumption.

In thermal models, particle velocities are determined by the local
temperature, collective flow velocity and the particle masses. For
relativistic particles, the thermal velocity along the beam axis is
a function of the particle's transverse mass, and therefore is
affected by transverse expansion. Lower freeze-out temperature
and/or larger transverse mass of higher $\pt$ particles are expected
to result in smaller thermal velocity in the longitudinal direction
and narrower BF~\cite{bf1,bf2}.

In this letter, we first give brief descriptions of the analysis
parameters and techniques. We then present measurements of the BF,
and its dependence on the width of the pseudorapidity window. We
test the boost-invariance of the BF, i.e. verify whether it is
independent of the position of same-width pseudorapidity windows. We
next examine the universality of the scaled BF. Finally, the scaling
property of the BF for particles within different transverse
momentum ranges is studied.

Our BF analysis is restricted to charged particles measured within
the STAR TPC detector~\cite{STARTPC}. This detector is well suited
for precise studies of correlation structures given its relatively
wide pseudorapidity range $-1.3 < \eta < 1.3$ and full azimuthal
acceptance. Recorded events were selected on the basis of a
minimum-bias trigger defined by the coincidence of two zero-degree
calorimeters (ZDCs) ~\cite{STARZDC} located at $\pm 18$~m from the
center of the TPC. Events are further required to have a primary
vertex position within 25 cm, longitudinally, of the TPC center and
within 1 cm, radially, of the beam line. This analysis is restricted
to charged particle tracks in the $\pt$ range $0.15<\pt<2.0$
GeV$/c$. After all these cuts, 5.7 million minimum-bias events were
selected for the analysis. Tracks are required to pass within 2 cm
of the primary vertex in order to reduce weak-decay contributions.
Tracks are further required to consist of a minimum of 15 measured
points and have a ratio of the numbers of measured to possible
points larger than 0.52 to avoid track splitting effects. These two
cuts minimize detector and track reconstruction effects, such as
ghost tracks, track splitting, and enable optimal momentum
resolution.

Figure 1(a) displays balance function obtained with five different
pseudorapidity windows, located at various positions, and with sizes
ranging from $|\ew| = 0.6$ to 2.6. It shows that the BF is strongly
dependent on the width of the pseudorapidity window. Vertical bars
shown in this and following figures indicate statistical errors
only. Statistical errors are smaller than the symbol sizes in
Fig.~1. Systematic errors are of the order of $5\%$ and due to
uncertainties in the track reconstruction efficiency associated with
the track cuts, and event-by-event variations of the vertex
position.

\begin{figure*}
\includegraphics[width=6.6in]{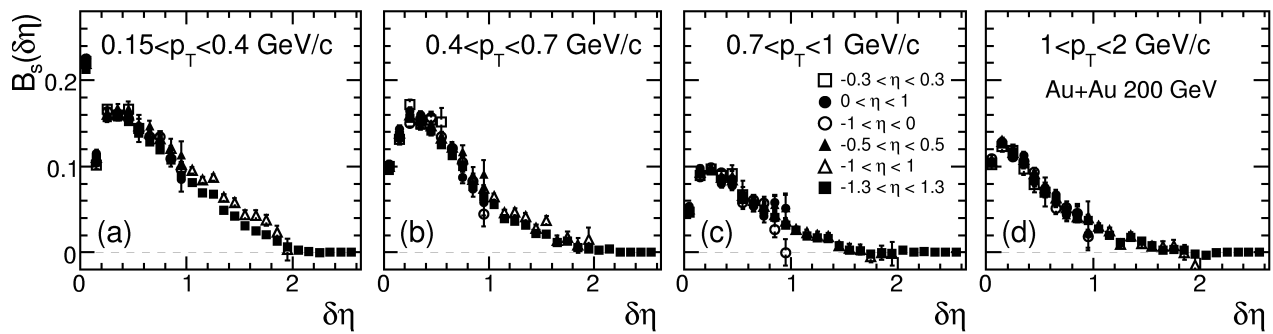}
\caption{\label{Fig. 2} $\bs(\de)$ based on $B(\de|\ew)$ values
measured in different pseudorapidity windows for particles in four
$\pt$ bins. The data are from $0-80\%$ Au+Au collisions at 200 GeV.
Error bars are statistical only. Systematic errors are of the order
of $5\%$.}
\end{figure*}

In order to test directly whether the BF is boost-invariant under
longitudinal translation within the STAR TPC, we examine, in
Fig.~1(b), five BFs measured in equal size ($|\ew| = 0.8$)
pseudorapidity windows located at different positions. One observes
that the five BFs overlap with one another thereby indicating that
the BF is independent of the position of the pseudorapidity window,
i.e., $B(\dt\eta|\eta_{\rm w})$ is invariant under a longitudinal
translation within the range $-1 < \eta < 1$. Note that the large BF
values measured at $\dt\eta=0.01$ arise in part from HBT and Coulomb
effects~\cite{bf2,star130}. We also considered five equal size and
non-overlapping windows, not shown in Fig.~1, and found similar
agreements.

In Fig.~1(c), we present scaled balance functions, $\bs$, calculated
with Eq.~2, obtained from BFs measured with four distinct
pseudorapidity window widths ($|\ew|= 0.6, 1, 2, 2.6$) and six
window positions. We find that the scaled balance functions have
equal shape and magnitude, and are identical within experimental
errors. Therefore $\bs$ is independent of the size and position of
the window $\ew$ in the pseudorapidity range $-1 < \eta < 1$. A
similar invariance of $\bs$ was observed in hadron-hadron
interactions over the whole rapidity range of produced
particles~\cite{na22bf}.

Lastly, we investigate whether the scaling property of the BF holds
for particles in different $\pt$ ranges and study how the width of
the BF changes with $\pt$. Figure 2 displays scaled balance function
obtained for four $\pt$ ranges: (0.15, 0.4), (0.4, 0.7), (0.7, 1)
and (1, 2) GeV$/c$, and the same pseudorapidity windows as used in
Fig.~1 (c). We find that the distributions measured in specific
$\pt$ intervals are independent of the size and position of the
pseudorapidity window used to carry out the measurement. We thus
conclude that the invariance of $\bs$ observed for $0.15<\pt<2.0$
GeV$/c$ also holds for small transverse momentum ranges.
\begin{table*}
\caption{\label{Table 1.}The widths $\la\delta\eta\ra$ of the
$\bs(\de)$ for four $\pt$ bins. The first and second errors are
statistical and systematic, respectively. The data are from $0-80\%$
Au+Au collisions at 200 GeV.}
\begin{tabular}{ccccccccc} \hline
 $\pt$(GeV$/c$) && (0.15, 0.4) && (0.4, 0.7) && (0.7, 1) && (1, 2) \\ \hline
$\la \delta\eta\ra$ && $0.652 \pm 0.006^{+0.081}_{-0.029}$
&& $0.609 \pm 0.008^{+0.049}_{-0.037}$&& $0.536 \pm0.016^{+0.047}_{-0.041}$
&& $0.487\pm 0.014^{+0.079}_{-0.021}$\\
\hline
\end{tabular}
\end{table*}

Comparing the distributions shown in Fig.~2(a) to 2(d), we observe
that the scaled balance function, $\bs(\de)$, changes significantly
in shape and amplitude with the transverse-momentum of final state
particles. The widths of $\bs(\de)$ defined as~\cite{star130} \beqar
\langle \dt \eta\rangle=\frac{\sum_i \bs(
\dt\eta_i)\dt\eta_i}{\sum_i \bs( \dt\eta_i)}\eeqar are presented in
Table I.  The first data point in Fig.~2(a) is affected by HBT
correlations, which result in a strong correlation at small relative
$\pt$. On the other hand, track merging effects deplete the balance
function at small $\delta\eta$. To assess the systematic
uncertainties on the extracted width, we use extrapolated values for
the data points at the two lowest $\delta\eta$ instead of their
measured ones in calculating the width. For the lower bound of
systematic uncertainty estimate, the extrapolations from the larger
$\delta\eta$ data are done by two functional forms in order to well
fit the data. One is exponential for the $\pt$ in (0.15, 0.4) GeV/c
and Gaussian for the other three $\pt$ bins. For the upper bound of
systematic uncertainty estimate, the extrapolated function is
multinomial for all four $\pt$ bins. Table I shows that the width of
the scaled BF becomes narrower for increasing $\pt$. This
observation is qualitatively consistent with expectations from
thermal models~\cite{bf2}.

\begin{figure*}
\includegraphics[width=6.6in]{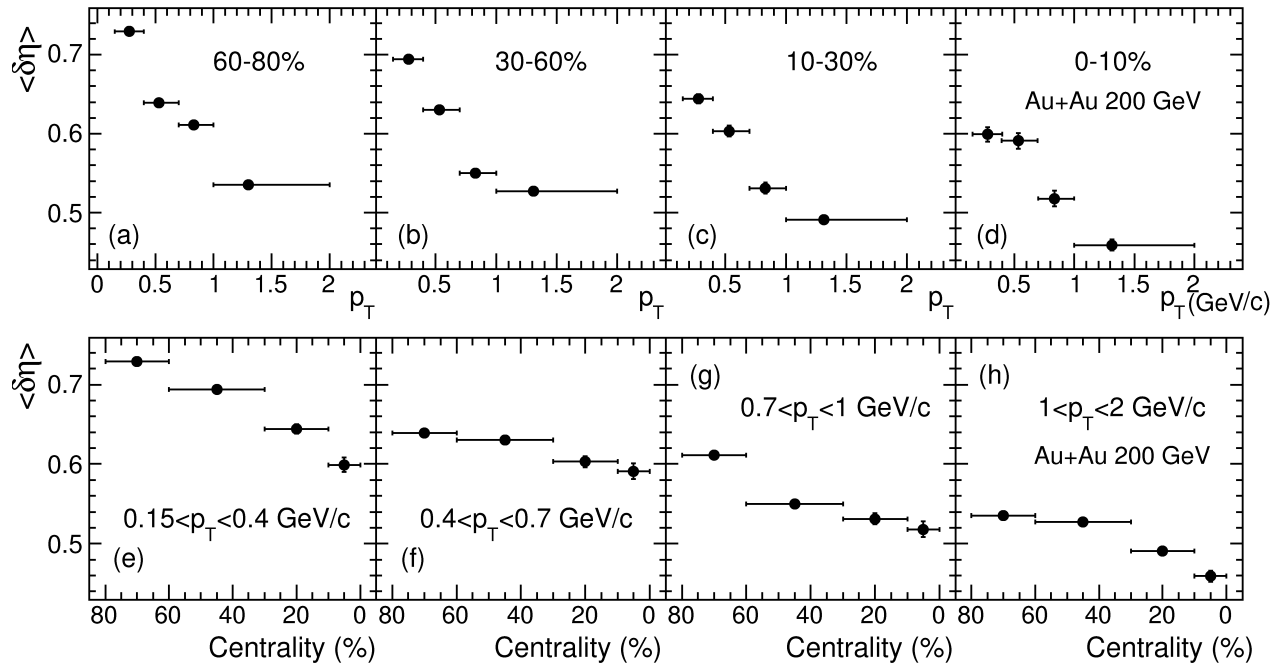}
\caption{\label{Fig. 3} Upper panel: the $\pt$ dependence of the
width of the BF in different centrality bins; Lower panel: the
centrality dependence of the width of the BF in different $\pt$
intervals. Data are from Au + Au collisions at 200 GeV. }
\end{figure*}

As shown in~\cite{star130}, the width of the BF decreases with
collision centrality. The decreases in the BF width with increasing
$\pt$ and increasing centrality could be associated with transverse
radial flow~\cite{radial}. In order to disentangle these effects, we
further study the $\pt$ dependence of $\langle\delta\eta\rangle$ in
different centrality bins. This is shown in the upper panel of
Fig.~3. It shows clearly that the width of the BF decreases with
increasing transverse momentum of final state particles in each
centrality bin. We also study the centrality dependence of
$\langle\delta\eta\rangle$ in different $\pt$ intervals. This is
presented in the lower panel of Fig.~3. It shows that the narrowing
of the BF with increasing centrality is present in all $\pt$ bins.
Our results demonstrate that the BF becomes narrower with increasing
$\pt$ in each given centrality bin, and in more central collisions
in each given $\pt$ bin. The width of BF depends on both centrality
and $\pt$. The origin of these narrowings and their possible
connections should provide more insight into the particle production
dynamics in relativistic heavy-ion collisions.

In summary, we present a first measurement of the longitudinal
scaling property of the charge balance function in Au + Au
collisions at 200 GeV with the STAR detector at RHIC. The results
demonstrate that within the pseudorapidity range $-1.3 < \eta <
1.3$, the balance function in equal size windows is independent of
the position of the window, and the balance function, when properly
scaled by the width of the pseudorapidity window, is found to be
independent of the position and size of the window. This scaling
property of the balance function is also observed for particles in
different $\pt$ ranges. It is further shown that the width of the
scaled BF decreases with increases of both the particle transverse
momentum and the collision centrality. We conclude that the scaling
property of the BF, observed in hadron-hadron
collisions~\cite{na22bf}, is also present in nucleus-nucleus
collisions at mid-rapidity at RHIC. This longitudinal property of BF
provides a good test for the hadronization mechanism in currently
available models~\cite{MC-BF}. The narrowing of the BF with
increasing $\pt$ and centrality warrants further investigation.

\vskip 0.5cm We thank Dr. Xin-Nian Wang for discussions on the
scaling behavior of the charge balance function. We thank the RHIC
Operations Group and RCF at BNL, the NERSC Center at LBNL and the
Open Science Grid consortium for providing resources and support.
This work was supported in part by the Offices of NP and HEP within
the U.S. DOE Office of Science, the U.S. NSF, the Sloan Foundation,
the DFG cluster of excellence `Origin and Structure of the Universe'
of Germany, CNRS/IN2P3, STFC and EPSRC of the United Kingdom, FAPESP
CNPq of Brazil, Ministry of Ed. and Sci. of the Russian Federation,
NNSFC, CAS, MoST, and MoE of China, GA and MSMT of the Czech
Republic, FOM and NWO of the Netherlands, DAE, DST, and CSIR of
India, Polish Ministry of Sci. and Higher Ed., Korea Research
Foundation, Ministry of Sci., Ed. and Sports of the Rep. Of Croatia,
Russian Ministry of Sci. and Tech, and RosAtom of Russia.

\end{document}
%